\documentclass[letterpaper, 10pt, conference]{ieeeconf}
\IEEEoverridecommandlockouts                              

\usepackage{times}

\usepackage{setspace}
\usepackage{url}
\spacing{1}

\usepackage{graphicx}		
\usepackage{wrapfig}
\usepackage[format=plain,font=footnotesize,labelfont=bf,labelsep=period]{caption}
\usepackage{sidecap} 
\usepackage{subfig}
\usepackage[export]{adjustbox}

\usepackage{amsmath}
\usepackage{amssymb}
\usepackage{mathtools}

\usepackage{pdfsync}
\usepackage[normalem]{ulem}
\usepackage{paralist}	
\usepackage[space]{grffile} 

\usepackage{color}

\usepackage{amsthm}

\newtheorem{theorem}{Theorem}
\newtheorem{corollary}{Corollary}
\newtheorem{lemma}{Lemma}
\theoremstyle{definition}
\newtheorem{definition}{Definition}
\theoremstyle{remark}
\newtheorem{remark}{Remark}
\theoremstyle{definition}

\theoremstyle{definition}

\newcommand{\R}{\mathbb{R}}

\newcommand{\gap}{\vspace{.1cm}}

\definecolor{darkblue}{RGB}{0,0,102}
\definecolor{lightblue}{RGB}{77,77,148}

\definecolor{gold}{RGB}{234, 170, 0}
\definecolor{metallic_gold}{RGB}{139, 111, 78}

\renewcommand{\cal}[1]{\mathcal{ #1 }}
\newcommand{\mb}[1]{\mathbf{ #1 }}
\newcommand{\bs}[1]{\boldsymbol{ #1 }}

\newcommand{\interior}{\mathrm{int}}

\newcommand{\intersect}{\cap}
\newcommand{\union}{\cup}

\DeclareMathOperator*{\argmin}{argmin}

\title{\textbf{Adaptive Safety with Control Barrier Functions}}
\author{Andrew J. Taylor and Aaron D. Ames
\thanks{Both authors are with the Department of Computing and Mathematical Sciences, California Institute of Technology, Pasadena, CA 91125, USA {\tt\small ajtaylor@caltech.edu, ames@caltech.edu}}
}

\begin{document}

\maketitle

\vspace{-.2cm}



\begin{abstract}
Adaptive Control Lyapunov Functions (aCLFs) were introduced 20 years ago, and provided a Lyapunov-based methodology for stabilizing systems with parameter uncertainty. The goal of this paper is to revisit this classic formulation in the context of safety-critical control. This will motivate a variant of aCLFs in the context of safety: \emph{adaptive Control Barrier Functions (aCBFs)}. Our proposed approach adaptively achieves safety by keeping the systems state within a safe set even in the presence of parametric model uncertainty. We unify aCLFs and aCBFs into a single control methodology for systems with uncertain parameters in the context of a Quadratic Program (QP) based framework. We validate the ability of this unified framework to achieve stability and safety in an adaptive cruise control (ACC) simulation.
\end{abstract}


\section{Introduction}


In many modern control applications, safety is of critical importance. It is impossible to model the system dynamics in these applications exactly---that is, parameters of the model may not match the real system. For instance, the mass and electrical properties of robotic systems are often approximate values. Thus, to truly enforce safety, it is necessary to quantify safety in the context of unknown parameters.

The use of Control Barrier Functions (CBFs) \cite{ames2014control, ames2017control} for ensuring safety of nonlinear control systems has become increasingly popular \cite{ nguyen2016exponential,wang2017safe, wang2018safe}. Controllers synthesized via CBFs rely on a model, and the guarantees they achieve may fail in the presence of model uncertainty. Robust control methods can ensure safety \cite{gurriet2018towards,xu2015robustness} or quantify how safety properties degrade \cite{kolathaya2018input} in the presence of model uncertainty, but may be overly conservative in restricting the behavior of the system.
Data-driven methods employing machine learning \cite{ohnishi2019barrier,cheng2019end} provide probabilistic safety guarantees, but may require episodic, offline training to improve model estimates \cite{fisac2018general}. 

In this paper, we focus on an online, \textit{adaptive} approach to ensuring that a system remains safe in the presence of model uncertainty. Adaptive control seeks to update a model of the system as it evolves to achieve stability or a desired level of performance \cite{miroslav1995nonlinear}. In particular, we build upon the idea of \textit{adaptive} Control Lyapunov Functions (aCLFs) \cite{krstic1995control}, which have been used to stabilize nonlinear systems in the presence of parametric model uncertainty \cite{krstic1996modular, li1997optimal, moore2014adaptive}. That is, the goal of this paper is to find conditions for adaptive safety (via Control Barrier Functions) equivalent to those derived for adaptive stability (via Control Lyapunov Functions). 

One challenge in developing adaptive control methods that guarantee safety is ensuring that the a nonlinear system's state remains within a prescribed safe set at all times. In contrast, guarantees on stability provided by aCLFs describe the behavior of the state and parameter estimation error jointly, allowing the state to grow large before stabilizing, as long as the parameter estimation error diminishes. To achieve this stricter guarantee of safety, we leverage stronger assumptions on the initial parameter estimation error. The end result are conditions for safety even under the presence of model uncertainty, i.e., that a system with unknown parameters can be rendered safe for all time. 

The main contribution of this paper is a formal methodology for ensuring safety in nonlinear (control affine) systems with parameter uncertainty through the formulation of {\it adaptive Control Barrier Functions (aCBFs)}. Like aCLFs, aCBFs provide a framework for updating model parameter estimates online, but do so to ensure safety. Unlike aCLFs, aCBFs require a different viewpoint on adaptive control to make stronger statements on the behavior of the system's state. To the best of our knowledge, our approach is the first that adaptively ensures safety utilizing CBFs. The definitions and results in this paper provide the first steps towards a framework for adaptive safety unifying both online and data-driven, episodic updates of model parameters.

This paper is organized as follows. Section II reviews CLFs and aCLFs and how quadratic program based controllers can be synthesized to adaptively stabilize a system. Section III discusses CBFs and how they can be used to ensure the safety of a system. Section IV provides the main result of the paper by defining aCBFs, and shows how a system can be rendered adaptively safe in the presence of model uncertainty. Section V offers a discussion on the assumptions and constraints made in the preceding section through a counter example. Section VI presents simulation results for an adaptive cruise control (ACC) system using both a safety-critical controller and a quadratic program based controller implementing an aCLF and an aCBF simultaneously.


\section{Adaptive Control Lyapunov Functions}
\label{sec:adaptiveCLF}


To develop provably correct controllers for nonlinear systems, it is typically assumed that the model is known.  Yet there are many practical applications where this assumption is not adequate. A simple illustration is a mechanical system whose parameters (masses, inertias, etc) are not completely known---and one may not want to treat the unknown model parameters as a perturbation from nominal parameters since this would only guarantee stability to a region corresponding to a bound on this difference (which also may not be known).  The purpose of this section, therefore, is to review the framework of adaptive Control Lyapunov Functions.

Consider a state space $\cal{X}\subset\R^n$ and a control input space $\cal{U}\subset\R^m$, where it is assumed $\cal{X}$ is path-connected and $\mb{0}\in\cal{X}$. Consider the affine dynamic system given by:
\begin{equation}
    \label{eqn:knowndynamics}
    \dot{\mb{x}} = \mb{f}(\mb{x})+\mb{g}(\mb{x})\mb{u}
\end{equation}
where $\mb{x}\in\cal{X}$, $\mb{u}\in\cal{U}$, $\mb{f}:\cal{X}\to\R^n$ and $\mb{g}:\cal{X}\to\R^{n\times m}$ are smooth on $\cal{X}$. We additionally assume $\mb{f}(\mb{0})=\mb{0}$. We will use the following definition, found in \cite{Khalil}, to study the stability of \eqref{eqn:knowndynamics}.

\begin{definition}[\it{Class $\cal{K}$ Function}]
A continuous function $\alpha:[0,a)\to\R_+$, with $a > 0$, is a \textit{class} $\cal{K}$ function ($\alpha\in\cal{K}$) if $\alpha(0)=0$ and $\alpha$ is strictly monotonically increasing. If $a=\infty$ and $\lim_{r\to\infty}\alpha(r)=\infty$, then $\alpha$ is said to be a \textit{class} $\cal{K}_\infty$ function ($\alpha\in\cal{K}_\infty$).
\end{definition}
Given this definition, we can define a Control Lyapunov Function (CLF) as in \cite{artstein1983stabilization}, \cite{lin1991universal}.
\begin{definition}[\it{Control Lyapunov Function (CLF)}]
A smooth function $V:\cal{X}\to\R_+$ is a \textit{Control Lyapunov Function} (CLF) for \eqref{eqn:knowndynamics} if there exists $\alpha_1,\alpha_2,\alpha_3\in\cal{K}_\infty$ such that:
\begin{align}
\alpha_1(\Vert\mb{x}\Vert)\leq V(\mb{x}) & \leq \alpha_2(\Vert\mb{x}\Vert) \\     
    \inf_{\mb{u}\in\cal{U}}\dot{V}(\mb{x},\mb{u})&\leq-\alpha_3(\Vert\mb{x}\Vert)
\end{align}
for all $x\in\cal{X}$.
\end{definition}
This definition can be constructed with $\alpha_1,\alpha_2,\alpha_3\in\cal{K}$, with resulting stability guarantees holding locally. The existence of a CLF for \eqref{eqn:knowndynamics} implies there exists a smooth (except at $\mb{x}=\mb{0}$) state-feedback controller $\mb{k}:\cal{X}\to\cal{U}$, that renders the origin globally asymptotically stable \cite{artstein1983stabilization}, \cite{sontag1989smooth}, noting that global refers to the state space $\cal{X}$.
$\mb{k}$ can be made continuous at $\mb{0}$ if $V$ satisfies the small control property \cite{sontag1989universal}.

Following the classic formulation of aCLFs in \cite{krstic1995control}, uncertainty in the dynamics \eqref{eqn:knowndynamics} appears as:
\begin{equation}
\label{eqn:uncertaindynamics}
\dot{\mb{x}}=\mb{f}(\mb{x})+\mb{F}(\mb{x})\bs{\theta}^{\star}+\mb{g}(\mb{x})\mb{u}, 
\end{equation}
where $\bs{\theta}^\star \in\Theta\subset\R^p$ is a vector of unknown parameters  and $\mb{F}:\cal{X}\to\R^{n\times p}$ is assumed to be smooth on $\cal{X}$ with $\mb{F}(\mb{0})=\mb{0}$. The impossibility of designing explicit controllers that are robust to unbounded unknown parameters suggests that we need to consider a larger class of controllers to stabilize \eqref{eqn:uncertaindynamics}. In particular, controllers that \emph{update} an estimate of the unknown parameters. These are called adaptive controllers, and take the form:
\begin{eqnarray}
\label{eqn:dynamiccontroller}
\mb{u}&=&\mb{k}(\mb{x},\widehat{\bs{\theta}})\\
\label{eqn:parameterupdate}
\dot{\widehat{\bs{\theta}}}&=&\bs{\Gamma} \bs{\tau}(\mb{x},\widehat{\bs{\theta}}),
\end{eqnarray}
where $\widehat{\bs{\theta}}\in\Theta$ represents an estimate of the parameters $\bs{\theta}^\star$ maintained by the controller, $\bs{\Gamma}\in\R^{n\times n}$ is a matrix adaptive gain, and $\bs{\tau}:\cal{X}\times\Theta\to\R^p$ is the adaptation law.  We make the following assumption on these functions:
\begin{description}
\item[(A1)] $\mb{k}$ is locally Lipschitz continuous on $(\cal{X}\backslash \{\mb{0}\})\times \Theta$ and $\mb{k}(\mb{0},\widehat{\bs{\theta}})=\mb{0}$,
\item[(A2)] $\bs{\tau}$ is locally Lipschitz continuous on $\cal{X}\times\Theta$,
\item[(A3)] $\bs{\Gamma}\in \R^{p\times p}$ is symmetric and positive-definite.
\end{description}
Introducing this parameter update results in a composite dynamic system:
\begin{equation}
    \label{eqn:compositedynamics}
    \begin{bmatrix}\dot{\mb{x}} \\ \dot{\widehat{\bs{\theta}}} \end{bmatrix} = \begin{bmatrix} \mb{f}(\mb{x})+\mb{F}(\mb{x})\bs{\theta}^\star+\mb{g}(\mb{x})\mb{k}(\mb{x},\widehat{\bs{\theta}}) \\ \bs{\Gamma}\bs{\tau}(\mb{x},\widehat{\bs{\theta}}) \end{bmatrix}
\end{equation}
Solutions to this system evolve in $\cal{X}\times\Theta$. Given this construction we introduce the following definition from \cite{krstic1995control}:

\begin{definition}[\it{Globally Adaptively Stabilizable}]
\label{def:adaptivestabilizable}
The system with unknown parameters \eqref{eqn:uncertaindynamics} is \textit{globally adaptively stabilizable} if there exists a dynamic controller of the form \eqref{eqn:dynamiccontroller}-\eqref{eqn:parameterupdate} satisfying (A1)-(A3) such that solutions $(\mb{x}(t),\widehat{\bs{\theta}}(t))$ of \eqref{eqn:compositedynamics} are globally bounded and $\lim_{t\to\infty} \mb{x}(t)=\mb{0}$.
\end{definition}

\begin{remark}
Note that the requirements for global adaptive stabilizability are rather weak in the sense that $\widehat{\bs{\theta}}$ is not required to converge to $\bs{\theta}^\star$. We will see, in fact, that convergence of $\widehat{\bs{\theta}}$ to $\bs{\theta}^\star$ is not necessary for $\mb{x}(t)$ to converge to the equilibrium.
\end{remark}

The strategy in designing adaptive controllers is to show that this problem is equivalent to a non-adaptive controller design problem. Such equivalence is shown via the notion of adaptive control Lyapunov functions as in \cite{krstic1995control}:

\begin{definition}[\it{Adaptive Control Lyapunov Function (aCLF)}]
\label{def:aCLF}
Let $\alpha_1(\cdot,\bs{\theta})$, $\alpha_2(\cdot,\bs{\theta})$, $\alpha_3(\cdot,\bs{\theta})\in\cal{K}_{\infty}$ for all $\bs{\theta}\in\Theta$. A smooth function $V_a:\cal{X}\times \Theta\to \R_+$, satisfying:
\begin{equation}
\alpha_{1}(\Vert \mb{x}\Vert,\bs{\theta})\le V_{a}(\mb{x},\bs{\theta})\le \alpha_{2}(\Vert \mb{x}\Vert,\bs{\theta}),
\end{equation}
is called an \textit{adaptive Control Lyapunov Function} (aCLF) for \eqref{eqn:uncertaindynamics} if there exists a symmetric positive-definite matrix $\bs{\Gamma}\in \R^{p\times p}$ such that for every $\bs{\theta}\in \Theta$, $V_a$ is a CLF for the system:
\begin{equation}
\label{eqn:modifiedsystem}
\dot{\mb{x}}=\mb{f}(\mb{x})+\mb{F}(\mb{x})\bs{\lambda}_{clf}(\mb{x},\bs{\theta}) +\mb{g}(\mb{x})\mb{u},
\end{equation}
where
\begin{eqnarray}
\label{eqn:lambdafunc}
\bs{\lambda}_{clf}(\mb{x},\bs{\theta}) \triangleq \bs{\theta}+\bs{\Gamma}\left(\frac{\partial V_{a}}{\partial \bs{\theta}}(\mb{x},\bs{\theta})\right)^T.
\end{eqnarray}
That is,
\begin{align}
\label{eqn:adaptiveCLFcondition}
\inf_{\mb{u}\in \cal{U}} & \left[ \frac{\partial V_{a}}{\partial \mb{x}}\left(\mb{f}(\mb{x}) + \mb{F}(\mb{x})\bs{\lambda}_{clf}(\mb{x},\bs{\theta})  +\mb{g}(\mb{x})\mb{u} \right) \right] \nonumber \\ & \hspace{4.5cm} \le -\alpha_3(\Vert \mb{x}\Vert, \bs{\theta}).
\end{align}
\end{definition}

Adaptive control Lyapunov functions can be used to obtain the following result establishing the equivalence between the original adaptive controller design problem and a non-adaptive one.

\begin{theorem}\label{thm:aCLF}{{\bf \cite{krstic1995control}}}
{\it System~\eqref{eqn:uncertaindynamics} is globally adaptively stabilizable iff there exists an aCLF for~\eqref{eqn:uncertaindynamics}.}
\end{theorem}

It is useful to give a sketch of the proof for the sufficiency portion of this result, as it will inform the proof of the analogous result in the context of control safety functions.

\begin{proof}[{\it Sketch}]
Assume that we have an aCLF $V_a$ for \eqref{eqn:uncertaindynamics}. As $V_a$ is a CLF for~\eqref{eqn:modifiedsystem} with $\bs{\theta} = \bs{\hat{\theta}}$, we can construct a smooth (away from $\mb{x}=\mb{0}$) controller $\mb{u}=\mb{k}(\mb{x},\widehat{\bs{\theta}})$ stabilizing~\eqref{eqn:modifiedsystem} (a specific example of a Lipschitz continuous controller will be given after the proof), i.e., we can construct a controller $\mb{u}=\mb{k}(\mb{x},\widehat{\bs{\theta}})$ such that:
\begin{equation}
\label{eqn:auxlyapunovineq}
L_{\tilde{\mb{f}}_{clf}}V_{a}(x,\widehat{\bs{\theta}})+L_{\mb{g}}V_{a}(\mb{x},\widehat{\bs{\theta}}) \mb{k}(\mb{x},\widehat{\bs{\theta}})\le -\alpha_3(\Vert \mb{x}\Vert,\widehat{\bs{\theta}}),
\end{equation}
where $\tilde{\mb{f}}$ is given by:
\begin{equation}
\label{eqn:ftilde}
\tilde{\mb{f}}_{clf}(\mb{x},\widehat{\bs{\theta}})=\mb{f}(\mb{x})+\mb{F}(\mb{x})\bs{\lambda}_{clf}(\mb{x},\widehat{\bs{\theta}}).
\end{equation}
We note that this controller only depends on the current estimate of the parameters $\widehat{\bs{\theta}}$, and does not depend on the actual parameters $\bs{\theta}^\star$. Define the parameter error:
\begin{equation}
\label{eqn:paramerror}
\widetilde{\bs{\theta}} = \bs{\theta}^\star-\widehat{\bs{\theta}}
\end{equation}
to be the difference between the actual and estimated parameters.
Consider now the candidate composite Lyapunov function:
\begin{equation}
\label{eqn:compositeV}
V(\mb{x},\widehat{\bs{\theta}})=V_a(\mb{x},\widehat{\bs{\theta}})+\frac{1}{2}\widetilde{\bs{\theta}}^T\bs{\Gamma}^{-1}\widetilde{\bs{\theta}}.
\end{equation}
Computing its derivative we obtain:
\begin{eqnarray}
\dot{V}& =& \dot{V}_a- \widetilde{\bs{\theta}}^T\bs{\Gamma}^{-1}\dot{\widehat{\bs{\theta}}}\nonumber\\
&\leq& -\alpha_3(\Vert \mb{x}\Vert,\widehat{\bs{\theta}})+ \widetilde{\bs{\theta}}^T\mb{a}(\mb{x},\widehat{\bs{\theta}})\nonumber \\
&& \hspace{.6 cm}-\left(\frac{\partial V_a}{\partial \bs{\theta}}(\mb{x},\widehat{\bs{\theta}})\right)\bs{\Gamma}\mb{a}(\mb{x},\widehat{\bs{\theta}}). \notag
\end{eqnarray}
where:
\begin{equation}
    \mb{a}(\mb{x},\widehat{\bs{\theta}}) = \left(\left(\frac{\partial V_{a}}{\partial \mb{x}}(\mb{x},\widehat{\bs{\theta}})\mb{F}(\mb{x})\right)^T   -\bs{\tau}(\mb{x},\widehat{\bs{\theta}})\right).
\end{equation}
It is now easy to see that using the update law
\begin{equation}
\label{eqn:clfupdatelaw}
\bs{\tau}(\mb{x},\widehat{\bs{\theta}})=\left(\frac{\partial V_a}{\partial \mb{x}}(\mb{x},\widehat{\bs{\theta}})\mb{F}(\mb{x})\right)^T
\end{equation}
implies
\begin{equation}
\dot{V}\le -\alpha_3(\Vert \mb{x}\Vert,\widehat{\bs{\theta}}),
\end{equation}
from which we conclude that the equilibrium point $(\mb{0},\bs{\theta}^\star)$ of \eqref{eqn:compositedynamics} is globally stable. In particular, we see that $(\mb{x}(t),\widehat{\bs{\theta}}(t))$ is globally bounded. It now follows from the LaSalle invariance principle that $\mb{x}(t)$ converges to the largest invariant subset of the collection of points $\mb{x}\in \cal{X}$ satisfying $\alpha_3(\Vert \mb{x}\Vert,\widehat{\bs{\theta}})=0$ which is the singleton $\mb{x}=\mb{0}$.
\end{proof} 

As noted in the preceding proof, given an aCLF $V_{a}$, we can correspondingly synthesize a Lipschitz continuous controller $\mb{u} = \mb{k}(\mb{x},\widehat{\bs{\theta}})$. This can be achieved in a point-wise optimal fashion by considering an optimization based control framework. In particular, since the aCLF condition \eqref{eqn:adaptiveCLFcondition} is satisfied, we can consider the following quadratic program:
\begin{align}
\label{eqn:aCLF_QP}
\tag{aCLF-QP}
\MoveEqLeft[7] \mb{k}(\mb{x},\widehat{\bs{\theta}}) =  \,\,\underset{\mb{u} \in \cal{U}}{\argmin}   \quad \frac{1}{2} \| \mb{u} \|^2  \\
\mathrm{s.t.} \quad \frac{\partial V_{a}}{\partial \mb{x}}&(\mb{x},\widehat{\bs{\theta}})\left(\tilde{\mb{f}}_{clf}(\mb{x},\widehat{\bs{\theta}}) +\mb{g}(\mb{x})\mb{u} \right)
\leq -\alpha_3(\Vert \mb{x}\Vert,\widehat{\bs{\theta}}) \nonumber
\end{align}
This QP based controller will be guaranteed to have a solution, again because \eqref{eqn:adaptiveCLFcondition} is satisfied, and is Lipschitz continuous \cite{morris2013sufficient}. Moreover, a closed form solution to this optimization problem, termed the {\it min-norm controller}, can be obtained via the KKT conditions \cite{boyd2004convex}. To see this, define:
\begin{align}
\phi_0(\mb{x},\widehat{\bs{\theta}}) ~\triangleq~ &  \frac{\partial V_{a}}{\partial \mb{x}}(\mb{x},\widehat{\bs{\theta}})\tilde{\mb{f}}_{clf}(\mb{x},\widehat{\bs{\theta}}) + \alpha_3(\Vert \mb{x}\Vert,\widehat{\bs{\theta}})  \nonumber\\
\bs{\phi}_1^T(\mb{x},\widehat{\bs{\theta}})  ~\triangleq~ &  \frac{\partial V_{a}}{\partial \mb{x}}(\mb{x},\widehat{\bs{\theta}}) \mb{g}(\mb{x})  \nonumber
\end{align}
wherein the solution to \eqref{eqn:aCLF_QP} follows from :
\begin{eqnarray}
\label{eqn:minnormgen}
\mb{k}(\mb{x},\widehat{\bs{\theta}}) & = &  \left\{  \begin{array}{lcr}
- \frac{\phi_0(\mb{x},\widehat{\bs{\theta}})\bs{\phi}_1(\mb{x},\widehat{\bs{\theta}})}{\bs{\phi}_1^T(\mb{x},\widehat{\bs{\theta}})\bs{\phi}_1(\mb{x},\widehat{\bs{\theta}})} & \mathrm{if} & \phi_0(\mb{x},\widehat{\bs{\theta}}) > 0 \\
0 & \mathrm{if} & \phi_0(\mb{x},\widehat{\bs{\theta}}) \leq 0
\end{array}
\right. \nonumber
\end{eqnarray}

\section{Control Barrier Functions}
The goal of this work is to provably enforce safety, even in the context of uncertain models. As a result, we will leverage the framework of Control Barrier Functions (CBFs) \cite{ames2014control}, \cite{ames2017control}, \cite{xu2015robustness}. This section, therefore, will review the basic concepts related to these functions and corresponding controller synthesis.

In the context of safety, we consider a set $S$ defined as the 0-superlevel set of a continuously differentiable function $h:\cal{X}\to\R$, yielding:
\begin{eqnarray}
\label{eqn:Sdef}
S &\triangleq& \{ \mb{x} \in \cal{X} ~\vert~ h(\mb{x}) \geq 0\}, \\
\label{eqn:Sboundarydef}
\partial S &\triangleq& \{ \mb{x} \in \cal{X} ~\vert~ h(\mb{x}) = 0\}, \\
\label{eqn:Sintdef}
\interior(S) &\triangleq& \{ \mb{x} \in \cal{X} ~\vert~ h(\mb{x}) > 0\},
\end{eqnarray}
We refer to $S$ as the \textit{safe set}.

Consider again the known dynamics \eqref{eqn:knowndynamics}. A feedback controller $\mb{u} = \mb{k}(\mb{x})$ induces closed loop dynamics:
\begin{eqnarray}
\label{eqn:certainclosedloop}
\dot{\mb{x}} = \mb{f}_{cl}(\mb{x}) \triangleq \mb{f}(\mb{x}) + \mb{g}(\mb{x}) \mb{k}(\mb{x})
\end{eqnarray}
which is assumed to be locally Lipschitz continuous. This assumption implies that for any initial condition $\mb{x}_0 \in \cal{X}$ there exists a maximum interval of existence $I(\mb{x}_0) = [0, \tau_{max})$ such that $\mb{x}(t)$ is the unique solution to \eqref{eqn:certainclosedloop} on $I(\mb{x}_0)$; in the case when $\mb{f}_{cl}$ is forward complete, $\tau_{max} = \infty$. This notation allows us to define forward invariance and safety:

\begin{definition}[\it{Forward Invariant}]
The set $S$ is \textit{forward invariant} if for every $\mb{x}_0 \in S$, $\mb{x}(t) \in S$ for $\mb{x}(0) = \mb{x}_0$ and all $t \in I(\mb{x}_0)$.
\end{definition}

\begin{definition}[\it{Safety}]
The system \eqref{eqn:certainclosedloop} is \textit{safe} with respect to the set $S$ if the set $S$ is forward invariant.
\end{definition}

It is desirable to achieve safety without the need to specify a specific controller as was done in \eqref{eqn:certainclosedloop}. This leads to the notion of Control Barrier Functions. Before defining these, we require the following definition as in \cite{ames2017control}:
\begin{definition}[\it{Extended Class $\cal{K}$ Function}]
A continuous function $\alpha:(-b,a)\to\R$, with $a,b > 0$, is an \textit{extended class} $\cal{K}$ function ($\alpha\in\cal{K}_e$) if $\alpha(0)=0$ and $\alpha$ is strictly monotonically increasing. If $a,b=\infty$, $\lim_{r\to\infty}\alpha(r)=\infty$, $\lim_{r\to-\infty}\alpha(r)=-\infty$. then $\alpha$ is said to be an \textit{extended class} $\cal{K}_{\infty}$ function ($\alpha\in\cal{K}_{\infty,e}$).
\end{definition}
This enables the following definition as in \cite{ames2017control}:

\begin{definition}[\it{Control Barrier Function (CBF)}]
\label{def:csf}
Let $S \subset \cal{X}$ be the 0-superlevel set of a continuously differentiable function $h:\cal{X}\to\R$. $h$ is a \textit{Control Barrier Function} (CBF) for $S$ if there exists an extended class $\cal{K}_\infty$ function $\alpha$ such that for the system \eqref{eqn:knowndynamics}:
\begin{align}
\label{eqn:csf_def}
\sup_{\mb{u} \in \cal{U}}  \left[ L_\mb{f} h(\mb{x}) + L_\mb{g} h(\mb{x}) \mb{u} +  \alpha(h(\mb{x}))  \right] \geq 0.
\end{align}
for all $ \mb{x} \in S $.
\end{definition}

We can consider the pointwise set consisting of all control values that render $S$ safe:
\begin{align}
\label{eqn:CBFpart1:safecontrollers}
K_{cbf}(\mb{x}) = \{ \mb{u} \in \cal{U} : L_\mb{f} h(\mb{x}) + L_\mb{g} h(\mb{x}) \mb{u} + \alpha(h(\mb{x})) \geq 0\}.
\end{align}
The main results of \cite{ames2014control,xu2015robustness} is that the existence of a CBF for $S$ implies the system \eqref{eqn:knowndynamics} can be rendered safe with respect to $S$:

\begin{theorem}
\label{thm:csf}
{\it Given a set $S \subset \cal{X}$ defined as the 0-superlevel set of continuously differentiable function $h: \cal{X} \to \R$, if $h$ is a CBF on $S$, then any Lipschitz continuous controller $\mb{k}$ such that $\mb{k}(\mb{x}) \in K_{cbf}(\mb{x})$ for all $\mb{x}\in S$ renders the system \eqref{eqn:knowndynamics} safe with respect to the set $S$.}
\end{theorem}
In addition, if $\mb{k}(\mb{x})\in K_{cbf}(\mb{x})$ for all $\mb{x}\in\cal{X}$, then the set $S$ is asymptotically stable in $\cal{X}$.


\section{Adaptive Control Barrier Functions}


Motivated by the construction of adaptive control Lyapunov functions (aCLFs), we now explore the notion of an adaptive Control Barrier Function. 

We again assume the control system has the form given in \eqref{eqn:uncertaindynamics}, wherein $\bs{\theta}^\star$ is a set of unknown parameters, and extend the previous construction of the safe set $S$ to be parameter dependent. In this case, we construct a family of safe sets parameterized by $\bs{\theta}$ and defined as the 0-superlevel sets of a continuously differentiable function $h_a:\cal{X}\times\Theta\to\R$: 
\begin{eqnarray}
\label{eqn:Sparamdef}
S_{\bs{\theta}} &\triangleq& \{ \mb{x} \in \cal{X} ~\vert~ h_a(\mb{x}, \bs{\theta}) \geq 0\}, \\
\label{eqn:Sparamboundarydef}
\partial S_{\bs{\theta}} &\triangleq& \{ \mb{x} \in \cal{X} ~\vert~ h_a(\mb{x},\bs{\theta}) = 0\}, \\
\label{eqn:Sparamintdef}
\interior(S_{\bs{\theta}}) &\triangleq& \{ \mb{x} \in \cal{X} ~\vert~ h_a(\mb{x},\bs{\theta}) > 0\},
\end{eqnarray}
In particular, we will see this construction allows the states in the state space that are considered safe to change according to the current estimate of the parameters. If set in the state space to be kept safe is independent of the parameters, the preceding construction is identical to that in \eqref{eqn:Sdef}-\eqref{eqn:Sintdef}.

Given this construction, we can define adaptively safe in a similar fashion to the definition of global adaptively stabilizable given in Definition \ref{def:adaptivestabilizable} (note that in this case we opt for a local rather than global definition).

\begin{definition}[\it{Adaptively Safe}]
\label{def:adaptivesafety}
The system with unknown parameters~\eqref{eqn:uncertaindynamics} can be rendered \textit{adaptively safe} with respect to a family of sets $S_{\widehat{\bs{\theta}}}$ if there exists a dynamic controller of the form~\eqref{eqn:dynamiccontroller}-\eqref{eqn:parameterupdate} satisfying (A1)-(A3) such that solutions $(\mb{x}(t),\widehat{\bs{\theta}}(t))$ of~\eqref{eqn:compositedynamics} controlled by~\eqref{eqn:dynamiccontroller}-\eqref{eqn:parameterupdate} satisfy $\mb{x}(t) \in S_{\widehat{\bs{\theta}}(t)}$ for all $t \in I(\mb{x}(0),\widehat{\bs{\theta}}(0))$.
\end{definition}
This definition implies that the state of the system must remain within a potentially time-varying set, $S_{\widehat{\bs{\theta}}(t)}$, even in the presence of uncertainty in the dynamics. It is not necessary that the parameters converge, or even that they remain bounded, as in the adaptively stabilizable formulation. As will be seen, this is inherently connected to the fact that safety does not force the system to converge to an equilibrium point, but only requires it remains within a set. 

Before defining aCBFs, we also specify that a set of adaptive gains $G$ is defined such that:
\begin{equation}
        \bs{\Gamma}\in G \implies \bs{\Gamma}~\mathrm{ satisfies~A(3).}
\end{equation}
We note that $G$ need not be all values of $\bs{\Gamma}$ satisfying A(3). We can now define aCBFs as an extension of Definitions \ref{def:aCLF} and \ref{def:csf}.

\begin{definition}[\it{Adaptive Control Barrier Function (aCBF)}]
\label{def:aCSF}
Let $S_{\bs{\theta}} \subset \cal{X}$ be a family of 0-superlevel sets of a continuously differentiable function $h_{a}:\cal{X}\times\Theta\to \R$, with $\frac{\partial h_a}{\partial\mb{x}}$ Lipchitz continuous. Then $h_{a}$ is an \textit{adaptive control barrier function} (aCBF) on the family of sets $S_{\bs{\theta}}$ over adaptive gains $G$ for~\eqref{eqn:uncertaindynamics} if for any $\bs{\theta}\in\Theta$ and $\bs{\Gamma}\in G$:
\begin{align}
\label{eqn:adaptiveCSFcondition}
\sup_{u\in \cal{U}}\left[\frac{\partial h_{a}}{\partial \mb{x}}(\mb{x},\bs{\theta})\left(\mb{f}(\mb{x}) + \mb{F}(\mb{x})\bs{\lambda}_{cbf}(\mb{x},\bs{\theta}) +\mb{g}(\mb{x})\mb{u} \right) \right] \geq 0.
\end{align}
with 
\begin{equation}
\label{eqn:lambdafunccsf}
\bs{\lambda}_{cbf}(\mb{x},\bs{\theta}) \triangleq \bs{\theta}-\bs{\Gamma}\left(\frac{\partial h_{a}}{\partial \bs{\theta}}(\mb{x},\bs{\theta})\right)^T.
\end{equation}
\end{definition}
Let us make a few observations of this definition:

\begin{remark}
As will be seen in the proof that an aCBF can ensure a system is adaptively safe, there is a requirement on the smallest eigenvalue of $\bs{\Gamma}$. As not every value of $\bs{\Gamma}$ satisfying A(3) will satisfy this requirement, we must consider a restricted set of values for $\bs{\Gamma}$, given by $G$. This leads to the incorporation of the set $G$ in the definition of an aCBF. If the family of sets $S_{\bs{\theta}}$ does not depend on $\bs{\theta}$, such that:
\begin{equation}
    \frac{\partial h_a}{\partial\bs{\theta}}(\mb{x},\bs{\theta}) \equiv \mb{0},
\end{equation}
then $\bs{\Gamma}$ will not appear in \eqref{eqn:adaptiveCSFcondition}. This implies $h_a$ being an aCBF for \eqref{eqn:uncertaindynamics} will not depend on $G$.
\end{remark}

\begin{remark}
The constraint in \eqref{eqn:adaptiveCSFcondition} differs from \eqref{eqn:csf_def} in that the term $\alpha(h_a(\mb{x},\bs{\theta}))$ does not appear. Rather, this closely resembles early definitions of \textit{barrier certificates} and \textit{Lyapunov barrier functions} \cite{prajna2005optimization}, \cite{wieland2007constructive}, \cite{tee2009barrier}, which did not allow the state to approach the boundary of the safe sets, enforcing forward invariance of level sets of $h_a$. As will be shown in Section \ref{sec:cex}, using the constraint from \eqref{eqn:csf_def} doesn't lead to the state safe set remaining forward invariant.
\end{remark}

We note that a QP-based Lipschitz continuous controller attaining safety can be constructed similarly to the \eqref{eqn:aCLF_QP} given an aCBF. We now have the necessary framework in which to present the main result of this paper---that the existence of an aCBF implies safety of the family of sets $S_{\bs{\widehat{\theta}}}$ even under parameter uncertainty.

\begin{theorem}
{\it Let $h_a:\cal{X}\to\R$ be an adaptive control barrier function on the family of sets $S_{\bs{\widehat{\theta}}}$ over $G$. Assume that $\widetilde{\bs{\theta}}_0 = \widetilde{\bs{\theta}}(0)$ with $\Vert \widetilde{\bs{\theta}}_0\Vert_2\leq c$ for $c > 0$ and $\mb{x}_0=\mb{x}(0)\in\interior(S_{\bs{\widehat{\theta}}_0})$ If there exists a positive definite gain matrix, $\bs{\Gamma}\in G$, such that:
\begin{equation}
\lambda_{min}(\bs{\Gamma}) \geq \frac{c^2}{2h_a(\mb{x}_0,\widehat{\bs{\theta}}_0)},
\end{equation} 
then there exists a Lipschitz continuous function $\bs{\tau}(\mb{x},\widehat{\bs{\theta}})$ such that for the update law:
\begin{equation}
    \dot{\widehat{\bs{\theta}}} = \bs{\Gamma}\bs{\tau}(\mb{x},\widehat{\bs{\theta}}),
\end{equation}
the family of sets $S_{\widehat{\bs{\theta}}}$ is forward invariant.}
\end{theorem}

\begin{table*}[ht]
\hrulefill
{\small
\begin{eqnarray}
\dot{h}(\mb{x},\widehat{\bs{\theta}},\mb{u})& =& \frac{\partial h_a}{\partial\mb{x}}(\mb{x},\widehat{\bs{\theta}})\left(\mb{f}(\mb{x})+\mb{F}(\mb{x})\bs{\theta}^\star+\mb{g}(\mb{x})\mb{u}\right)+\frac{\partial h_a}{\partial\bs{\theta}}(\mb{x},\widehat{\bs{\theta}})\dot{\widehat{\bs{\theta}}}+\widetilde{\bs{\theta}}^\top\bs{\Gamma}^{-1}\dot{\widehat{\bs{\theta}}} \nonumber\\
&=& \frac{\partial h_{a}}{\partial \mb{x}}(\mb{x},\widehat{\bs{\theta}})\left(\mb{f}(\mb{x})+\mb{F}(\mb{x})\bs{\theta}^\star+\mb{g}(\mb{x})\mb{u}\right) +\frac{\partial h_{a}}{\partial \bs{\theta}}(\mb{x},\widehat{\bs{\theta}})\bs{\Gamma}\bs{\tau}(\mb{x},\widehat{\bs{\theta}})+\frac{\partial h_a}{\partial\mb{x}}(\mb{x},\widehat{\bs{\theta}})\mb{F}(\mb{x})\left(\widehat{\bs{\theta}}-\bs{\Gamma}\left(\frac{\partial h_a}{\partial \bs{\theta}}(\mb{x},\widehat{\bs{\theta}})\right)^\top\right)\nonumber\\& &-\frac{\partial h_a}{\partial\mb{x}}(\mb{x},\widehat{\bs{\theta}})\mb{F}(\mb{x})\left(\widehat{\bs{\theta}}-\bs{\Gamma}\left(\frac{\partial h_a}{\partial \bs{\theta}}(\mb{x},\widehat{\bs{\theta}})\right)^\top\right)+\widetilde{\bs{\theta}}^T\bs{\tau}(\mb{x},\widehat{\bs{\theta}})\notag\\
&=& \frac{\partial h_a}{\partial\mb{x}}(\mb{x},\widehat{\bs{\theta}})\left(\mb{f}(\mb{x}) + \mb{F}(\mb{x})\left(\widehat{\bs{\theta}}-\bs{\Gamma}\left(\frac{\partial h_a}{\partial \bs{\theta}}(\mb{x},\widehat{\bs{\theta}})\right)^\top\right)+\mb{g}(\mb{x})\mb{u}\right) \nonumber \\ & &  + \frac{\partial h_a}{\partial\mb{x}}(\mb{x},\widehat{\bs{\theta}})\mb{F}(\mb{x})\left(\widetilde{\bs{\theta}}+\bs{\Gamma}\left(\frac{\partial h_a}{\partial \bs{\theta}}(\mb{x},\widehat{\bs{\theta}})\right)^\top\right)+\frac{\partial h_a}{\partial\bs{\theta}}(\mb{x},\widehat{\bs{\theta}})\bs{\Gamma}\bs{\tau}(\mb{x},{\widehat{\bs{\theta}}}) + \widetilde{\bs{\theta}}^\top\bs{\tau}(\mb{x},\widehat{\bs{\theta}})\nonumber \\ &\geq& \left(\frac{\partial h_a}{\partial\bs{\theta}}(\mb{x},\widehat{\bs{\theta}})\bs{\Gamma}+\widetilde{\bs{\theta}}^\top\right)\left(\left(\frac{\partial h_a}{\partial\mb{x}}(\mb{x},\widehat{\bs{\theta}})\mb{F}(\mb{x})\right)^\top+ \bs{\tau}(\mb{x},\widehat{\bs{\theta}})\right) \nonumber\\ &\geq& 0 \nonumber
\end{eqnarray}
}
\hrulefill
\caption{Calculation of $\dot{h}$ as used in the proof of the main result.}
\label{table:hdotcalc}
\end{table*}

The main idea is to approach the proof much in the same way as the proof of Theorem \ref{thm:aCLF}. Yet the construction of a composite CBF as was done in \eqref{eqn:compositeV} in the case of aCLFs requires more care. Adding the parameter error term would result in the composite safety function 0-superlevel set properly containing the 0-superlevel set of the aCBF, adding additional states to the set that can be rendered safe. This extension of the safe set can be quantified if the parameter estimates (and thus the parameter error) remains bounded, as in the case of aCLFs, but this is not guaranteed given the necessary form of $\bs{\tau}$. 
\vspace{-0.5cm}\begin{proof}
Define the following composite candidate CBF for the extended system dynamics \eqref{eqn:compositedynamics}:
\begin{eqnarray}
\label{eqn:compositeh}
h(\mb{x},\widehat{\bs{\theta}}) = h_a(\mb{x},\widehat{\bs{\theta}})- \frac{1}{2}\widetilde{\bs{\theta}}^T\mb{\Gamma}^{-1}\widetilde{\bs{\theta}}
\end{eqnarray}
By assumption, $\mb{x}_0\in\interior(S_{\widehat{\bs{\theta}}_0})$, implying that $h_a(\mb{x}_0,\widehat{\bs{\theta}}_0) > 0$. Further, our assumption that $\Vert\widetilde{\bs{\theta}}_0\Vert_2\leq c$ implies that:
\begin{equation}
    \frac{1}{2}\tilde{\bs{\theta}}_0^\top\bs{\Gamma}^{-1}\tilde{\bs{\theta}}_0 \leq \frac{1}{2\lambda_{min}(\bs{\Gamma})}\Vert \tilde{\bs{\theta}}_0\Vert^2_2 \leq \frac{c^2}{2\lambda_{min}(\bs{\Gamma})}
\end{equation}
Therefore, choosing $\bs{\Gamma}$ such that 
\begin{equation}
    \label{eqn:gainineq}
    \lambda_{min}(\bs{\Gamma}) \geq \frac{c^2}{2h_a(\mb{x}_0,\widehat{\bs{\theta}}_0)}
\end{equation}
leads to:
\begin{equation}
    \label{eqn:compositesafeic}
    h(\mb{x}_0, \widehat{\bs{\theta}}_0) \geq 0
\end{equation}
Now consider the time derivative of $h$ as given in Table \ref{table:hdotcalc}. The second equality follows the addition and subtraction of the term:
\begin{equation}
    \frac{\partial h_a}{\partial\mb{x}}(\mb{x},\widehat{\bs{\theta}})\mb{F}(\mb{x})\left(\widehat{\bs{\theta}}-\bs{\Gamma}\left(\frac{\partial h_a}{\partial \bs{\theta}}(\mb{x},\widehat{\bs{\theta}})\right)^\top\right)
\end{equation}
The third equality is a rearrangement revealing the form of the aCBF time derivative as given in \eqref{eqn:adaptiveCSFcondition}-\eqref{eqn:lambdafunccsf}. In particular, condition \eqref{eqn:adaptiveCSFcondition} permits the choice of an input $\mb{u}$ such that the first inequality is satisfied. Choosing the update law $\bs{\tau}$ as:
\begin{equation}
\label{eqn:csfupdatelaw}
  \bs{\tau}(\mb{x},\widehat{\bs{\theta}}) = -\left(\frac{\partial h_a}{\partial\mb{x}}(\mb{x},\widehat{\bs{\theta}})\mb{F}(\mb{x})\right)^\top
\end{equation}
results in the last inequality. This inequality, in conjunction with \eqref{eqn:compositesafeic} and the comparison lemma in \cite{Khalil} imply that 
\begin{equation}
    h(\mb{x}(t), \widehat{\bs{\theta}}(t)) \geq 0
\end{equation}
for all $t\geq 0$. Given the construction of $h$ in \eqref{eqn:compositeh}, it follows that:
\begin{equation}
    h_a(\mb{x}(t),\widehat{\bs{\theta}}(t))\geq \frac{1}{2}\widetilde{\bs{\theta}}(t)^\top\Gamma^{-1}\widetilde{\bs{\theta}}(t)\geq 0.
\end{equation}
Lastly, we conclude that $\mb{x}(t)\in S_{\widehat{\bs{\theta}}(t)}$ for $t\geq 0$.
\end{proof}

The proof reveals that superlevel sets of $h$ are forward invariant. As $h$ can not be computed without knowing the true parameters $\bs{\theta}^\star$, it is not possible to set $\dot{h}\geq-\alpha(h)$ as is typical with CBFs. Furthermore, we have that $h_a \geq h$, implying that $-\alpha(h_a)\leq-\alpha(h)$. Thus setting $\dot{h}\geq-\alpha(h_a)$ does not yield the desired lower bound on $\dot{h}$.
One may note that setting $\dot{h}\geq-\alpha(h_a)$ leads to $\dot{h}\geq 0$ when $h_a=0$, or when the state is on the boundary of the safe set. This fact is concurrent with the common forward invariance proof technique utilizing Nagumo's theorem \cite{nagumo1942lage}. Despite this, it is in fact possible to construct simple examples (in $\R^2$) such that the state must leave the safe set defined by $h_a$ for any choice of differentiable $\alpha$ and $\bs{\Gamma}$ as shown in Section \ref{sec:cex}.

\begin{remark}
The assumption on $\widetilde{\bs{\theta}}$ implies that the initial parameter error must be bounded, unlike the aCLF formulation. This is due to the fact that we seek to keep a particular set forward invariant. In contrast, the only set kept provably forward invariant in the aCLF formulation is the sublevel set of the composite Lyapunov function $V$ corresponding to the initial conditions $(\mb{x}(0),\widehat{\bs{\theta}}(0))$. Evaluating that set would too require assumptions on the boundedness of $\widetilde{\bs{\theta}}(0)$. Additionally, while this may seem restrictive, we note that the input for the system will not be chosen to be robust to all uncertainties in this initial uncertainty set. Rather, the uncertainty will be handled by adapting parameter estimates.
\end{remark}

\begin{remark}
The lower bound on the adaptive gain allows us to ensure that the system can adapt quickly enough to ensure safety from the given initial condition. Initial distance from the safety set boundary and smaller possible initial parameter error allow the adaptive gain to be made smaller.
\end{remark}

A quadratic program based controller similar to \eqref{eqn:aCLF_QP} can be constructed using an aCBF. To this end, we adopt the safety-critical control formulation in \cite{wang2017safe, gurriet2018towards} that filters a desired but potentially unsafe controller $\mb{k}_d:\cal{X}\times\Theta\to\cal{U}$ to find the nearest safe control action:
\begin{align}
\label{eqn:safety_critical_qp}
\tag{aCBF-QP}
 \hspace{-.7cm}\mb{k}(\mb{x},\widehat{\bs{\theta}}) =  \,\,\underset{\mb{u} \in \cal{U}}{\argmin} &   \quad \frac{1}{2} \| \mb{u}-\mb{k}_d(\mb{x},\widehat{\bs{\theta}}) \|^2  \\
  \mathrm{s.t.} \quad &\frac{\partial h_{a}}{\partial \mb{x}}(\mb{x},\widehat{\bs{\theta}}) (\tilde{\mb{f}}_{cbf}(\mb{x},\widehat{\bs{\theta}})+\mb{g}(\mb{x})\mb{u})
\geq 0 \nonumber
\end{align}
where $\widetilde{\mb{f}}_{cbf}$ is defined like $\widetilde{\mb{f}}_{clf}$ in \eqref{eqn:ftilde}. As with \eqref{eqn:aCLF_QP}, this quadratic program has a closed form solution.

\section{Analysis of aCBF Formulation}
\label{sec:cex}
In this section we analyze the aCBF conditions to verify that, in fact, they do not appear overly conservative.  In particular, changing the aCBF condition $\dot{h}_a \geq 0$ to $\dot{h}_a\geq-\alpha(h_a)$ does not necessarily lead to adaptive safety.
Consider the simple dynamic system given by: 
\begin{equation}
    \label{eqn:cexdynamics}
    \dot{x} = \theta + u
\end{equation}
with $\theta$ unknown and the safety function $h_a(x) = 1-x^2$ defining the state safe set $S = \{x\in\R~\vert~x^2\leq 1\}$. Assume that $x_0\in\interior(S)$ and $\tilde{\theta}^2_0\leq c^2$. The resulting composite safety function is given by:
\begin{equation}
    \label{eqn:cex2compsafe}
    h(x,\hat{\theta}) = h_a(x) - \frac{1}{2}\gamma^{-1}\tilde{\theta}^2
\end{equation}
with any $\gamma$ satisfying:
\begin{equation}
    \gamma \geq \frac{c^2}{2h_a(x_0)}.
\end{equation}
We additionally define the following sets:
\begin{align}
    \label{eqn:cexU}
    U = & \,\,\{(x,\tilde{\theta})\in\R^2~\vert~ x\in S\} \\
        \label{eqn:cexH0}
    H_0 = & 
    \,\,\{(x,\tilde{\theta})\in\R^2~\vert~ h(x,\hat{\theta})\geq 0\}
\end{align}
We note that the set $U$ extends infinitely along the $\tilde{\theta}$-axis, and completely contains $H_0$. Furthermore, $H_0\intersect\partial U = \{(-1,0),~(1, 0)\}$. The time derivative of the composite safety function is given by:
\begin{equation}
     \dot{h}(x,\hat{\theta},u) = -2x(\hat{\theta}+u) +\tilde{\theta}(-2x + \tau(x))
\end{equation}
for $\dot{\hat{\theta}}=\gamma\tau(x)$. Choosing the update law $\tau(x) = 2x$
and controller $u = -\hat{\theta}+\frac{1}{2}x\alpha(h_a(x))$, with extended $\cal{K}_\infty$ function $\alpha$, we have:
\begin{equation}
\dot{h}(x,\hat{\theta}) = -x^2\alpha(h_a(x)) \geq -\alpha(h_a(x)).
\end{equation}
as when $\alpha(h_a(x))\geq0$, $x^2\leq1$, and when $\alpha(h_a(x))\leq 0$, $x^2 \geq 1$. Noting the construction of $U$, we have the implication that $ (x,\tilde{\theta})\in U\implies \dot{h}(x,\hat{\theta}) \leq 0.$ The closed-loop state and parameter error dynamics are given by:
\begin{equation}
    \label{eqn:cntrex2cl}
    \begin{bmatrix}\dot{x} \\ \dot{\tilde{\theta}} \end{bmatrix} = \begin{bmatrix} \tilde{\theta} +\frac{1}{2}x\alpha(h_a(x)) \\ -2\gamma x \end{bmatrix} = \begin{bmatrix} \tilde{\theta}-F(x) \\ -g(x)\end{bmatrix},
\end{equation}
which has an unstable equilibrium point at the origin. This system is an example of a Li\'enard system (like the Van der Pol oscillator) as in \cite{perko2013differential}, with $F(x) = -\frac{1}{2}x\alpha(h_a(x))$ and $g(x) = 2\gamma x$. For systems of the this form, the following theorem, attributed to Li\'enard, provides the existence of a unique, stable limit cycle:

\begin{theorem}[Li\'enard's Theorem, \cite{perko2013differential}]
{\it Under the assumption that $F,g\in C^1(\R)$, $F$ and $g$ are odd functions of $x$, $xg(x)>0$ for $x\neq 0$, $F(0)=0$, $F'(0)<0$, $F$ has single positive zero at $x=a$, and $F$ increases monotonically to infinity for $x\geq a$ as $x\to\infty$, it follows that the Li\'enard system \eqref{eqn:cntrex2cl} has exactly one limit cycle and it is stable.}
\end{theorem}
If $\alpha$ is continuously differentiable in addition to an extended $\cal{K}_\infty$ function, the assumptions of this theorem are met by the functions given in \eqref{eqn:cntrex2cl}. We note that $a=1$ in this given example. Thus we can conclude that the system \eqref{eqn:cntrex2cl} has a stable periodic orbit, which we denote $\Phi$. We denote the open set in $\R^2$ enclosed by the limit cycle as $\interior(\Phi)$. Additionally, the proof of this theorem as in \cite{perko2013differential} implies the following corollary regarding the stable limit cycle:
\begin{corollary}
{\it The stable limit cycle $\Phi$ is symmetric about the origin and passes through a point, denoted as $P_2 = (x_2, \tilde{\theta}_2)$, such that $x_2 > a$.}
\end{corollary}
Given that $a=1$, this corollary reveals that the stable limit cycle leaves the set $U$, for which the state is considered safe. Additionally, as the limit cycle is symmetric about the origin, and the origin is an unstable equilibrium, the origin is contained in $\interior(\Phi)$.

This corollary also implies that $H_0\subset(\Phi\union\interior(\Phi))$. To see this, note that as the limit cycle encircles the origin, it must reenter the set $U$ after leaving the point $P_2$. At any point $v=(v_1,v_2)\in U$ that the limit cycle enters, we must have $h(v)\leq 0$, given the two points in $H_0\intersect\partial U$. Once the limit cycle enters $U$, we have $\dot{h}\leq 0$ until the limit cycle leaves $U$ as previously noted. Thus, $h\leq0$ along the portion of the limit cycle contained in $U$, implying $H_0\subset(\Phi\union\interior(\Phi))$. To complete the proof, we will employ the following definition and lemma from \cite{Khalil}:
\begin{definition}[Positive Limit Set]
The positive limit set $L^+$ is defined as all points $p\in\R^2$ such that there is a sequence $\{t_n\}$ with $t_n\to\infty$ as $n\to\infty$, and $(x(t_n),\tilde{\theta}(t_n))\to p$ as $n\to\infty$. 
\end{definition}

\begin{lemma}
{\it If a solution $(x(t),\tilde{\theta}(t))$ of \eqref{eqn:cntrex2cl} is bounded for $t\geq 0$, then its positive limit set $L^+$ is a nonempty, compact, invariant set, and $(x(t),\tilde{\theta}(t))$ approaches $L^+$ as $t\to\infty$.}
\end{lemma}

\begin{figure}[h]
    \hspace*{-0.6 cm}
    \centering
    \includegraphics[scale =0.43]{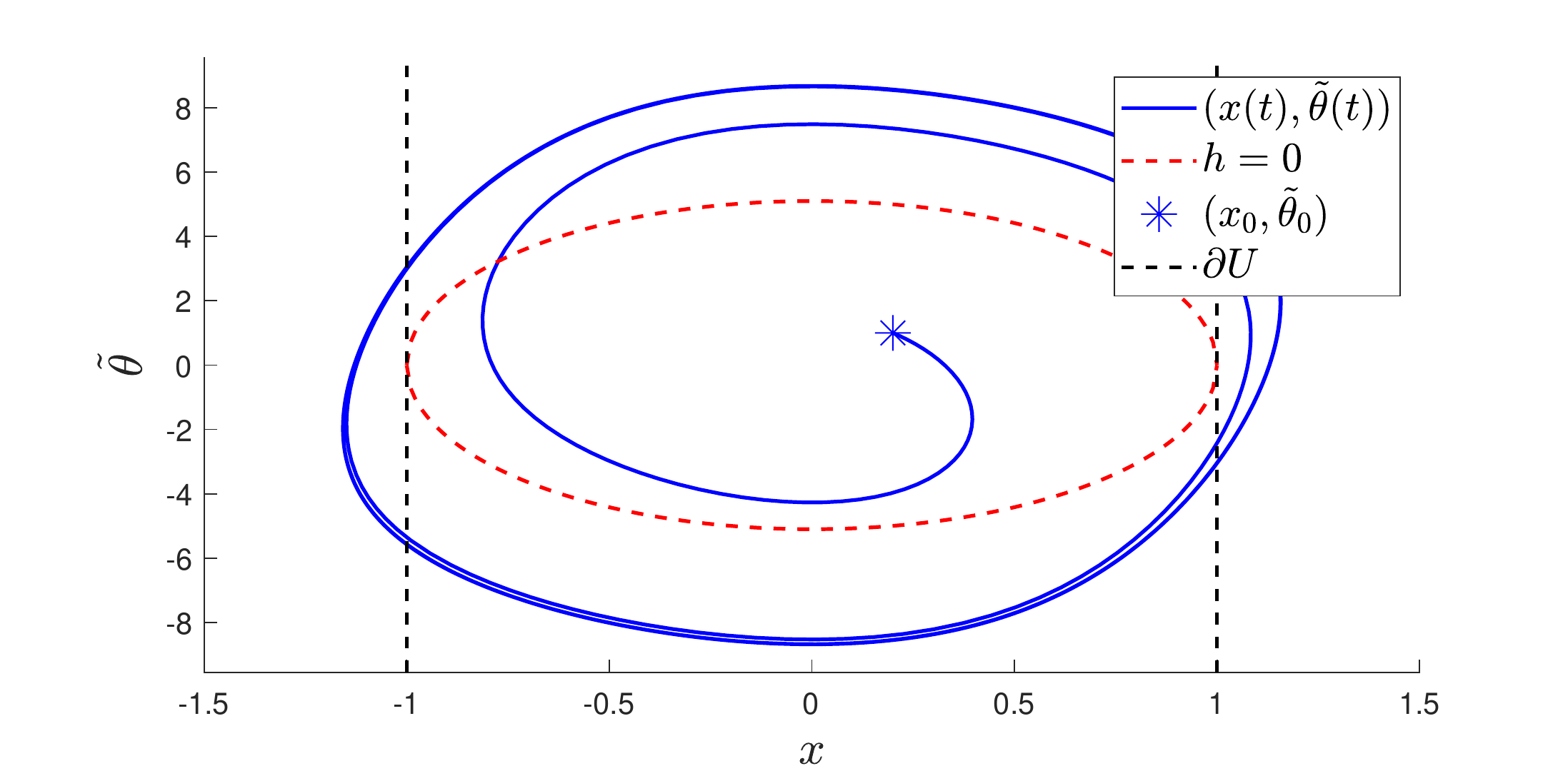}
    \caption{Evolution of the system governed by \eqref{eqn:cntrex2cl} with $\alpha(r) = kr$, $k = 10$, $(x_0,\tilde{\theta}_0) = (0.2, 1)$, $c = 5$, and $\gamma = 26$ achieving the lower bound.}
    \label{fig:cex2pp}
\end{figure}

We note that the unstable equilibrium point is not contained within the positive limit set $L^+$. As the 0-superlevel set of $h$, and thus all possible initial conditions given our bound on $\tilde{\theta}_0$, are contained inside the limit cycle, all solutions to \eqref{eqn:cntrex2cl} are bounded (by the limit cycle). Furthermore, $L^+=\Phi$, and thus all solutions starting in the 0-superlevel approach set approach $\Phi$. As the point $P_2\in\Phi$, and $P_2\notin U$, we see that any solution starting in the 0-superlevel set of $h$ leaves the desired state safe set $S$. Hence, the relaxation does not achieve safety of the state as desired, as seen in Figure \ref{fig:cex2pp}.



\section{Adaptive Cruise Control}
To demonstrate how an aCBF can be used to render a system adaptively safe, we consider the problem of adaptive cruise control (ACC) as posed in \cite{ames2014control}. The dynamics of the system are given by:
\begin{equation}
    \frac{\mathrm{d}}{\mathrm{d}t}\begin{bmatrix}v \\ D \end{bmatrix}= \begin{bmatrix} 0 \\ v_0 - v \end{bmatrix} -\frac{1}{m}\begin{bmatrix}1 & v & v^2 \\ 0 & 0 & 0 \end{bmatrix}\begin{bmatrix}f_0 \\ f_1\\ f_2 \end{bmatrix} +  \begin{bmatrix} \frac{1}{m} \\ 0 \end{bmatrix} u
\end{equation}
with $v$ the velocity of the vehicle, $D$ the distance between the vehicle and a leading vehicle traveling at a fixed velocity $v_0$, $m$ the vehicle's mass, and $f_0$, $f_1$, and $f_2$ unknown parameters associated with rolling frictional force. In this problem, we seek to drive the velocity to a desired velocity, $v_d$, while simultaneously ensuring the distance between the vehicles satisfies a safety constraint given by:
\begin{equation}
    D \geq 1.8 v.
\end{equation}
The parameters $f_0$, $f_1$, and $f_2$ are often determined empirically, and if they are not accurate, the desired velocity may not be accurately tracked. Furthermore, if the parameters do not exactly match the true parameters, it may not be possible to certify that the system will satisfy the safety constraint.

\begin{figure*}
    \hspace*{-0.5 cm}
     \centering
    \begin{subfloat}
        {\includegraphics[trim = {0, 0, 0, 24 cm}, clip ,scale =0.45, valign=t]{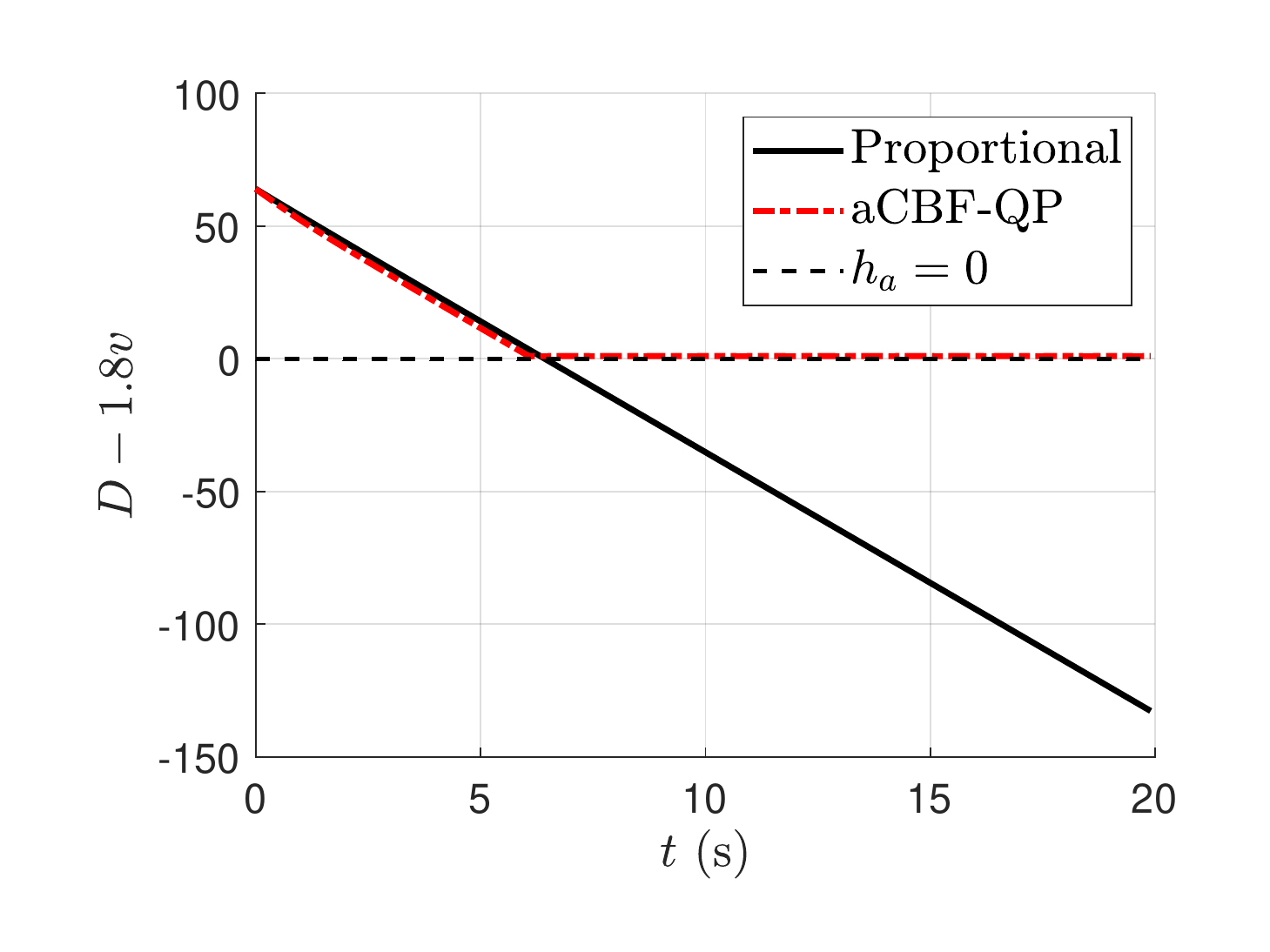}}
    \end{subfloat}
    \hfill
    \hspace*{-.5 cm}
    \begin{subfloat}
        {\includegraphics[scale =0.45, valign =t]{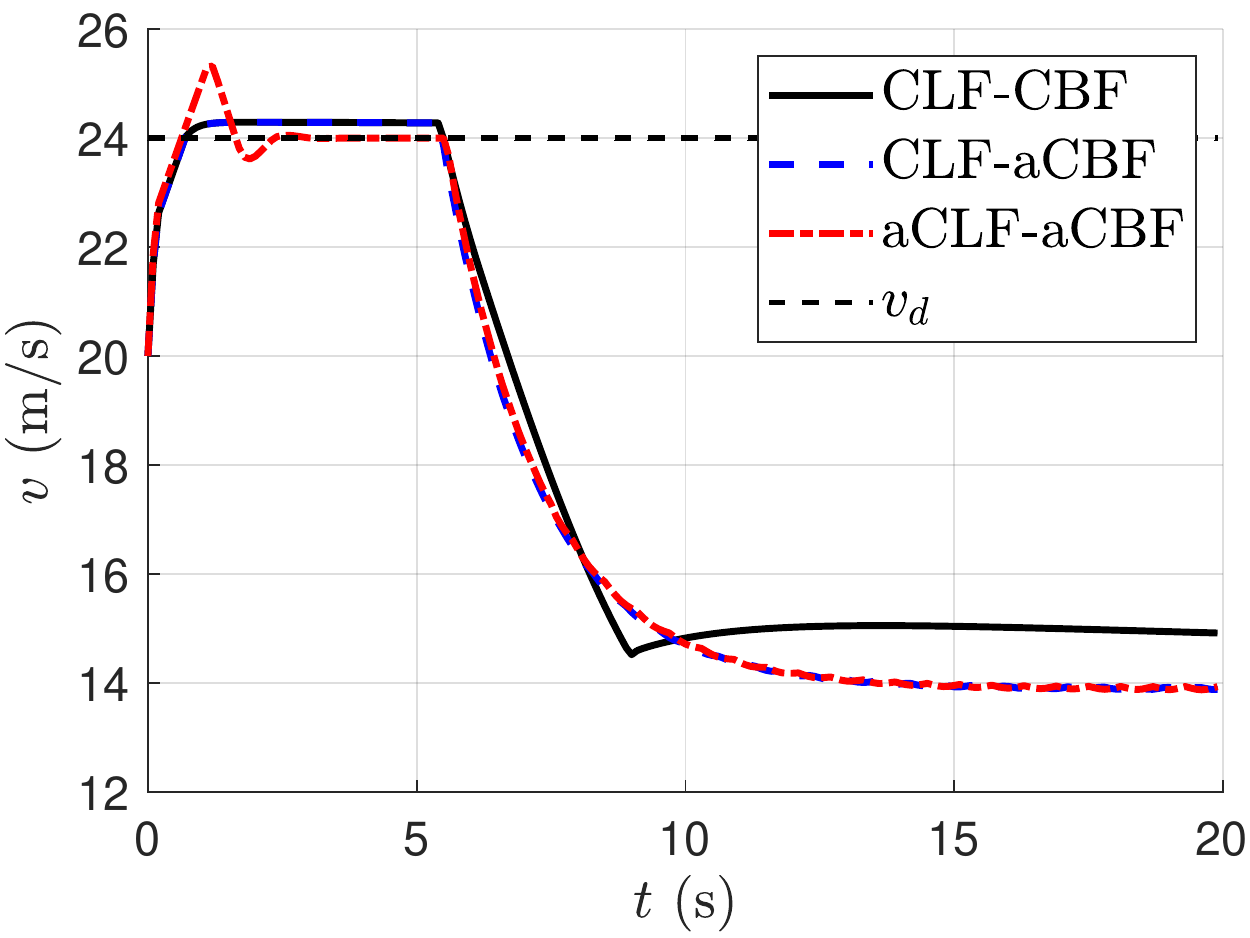}}
    \end{subfloat}
    \hfill
    \begin{subfloat}
      {\includegraphics[scale =0.45,valign=t]{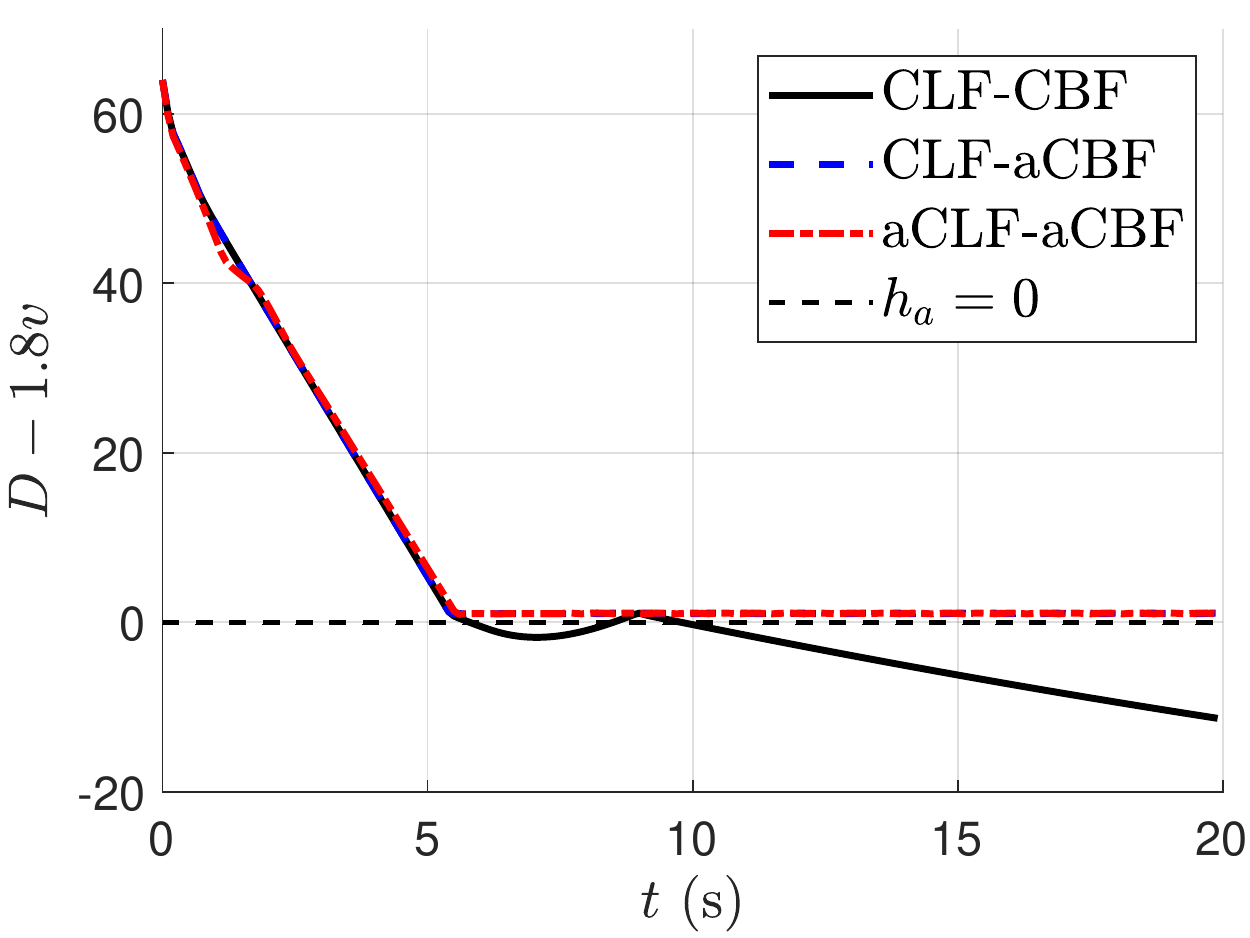}}
    \end{subfloat}
        \caption{Comparison of different adaptive and non-adaptive control methodologies. The aCBF-QP is able to enforce safety of the proportional controller (left). An aCLF controller is able to track the desired velocity with zero steady state error (center). Both aCBF controllers are able to keep the vehicle within the safe region for all time (right)}
    \label{fig:accsimresults}
\end{figure*}
    

The control objective of tracking a desired velocity can be achieved with a hand-designed controller $\mb{k}_d$ or encoded using a CLF, and the safety constraint can be encoded using a CBF. Additional constraints on the maximum acceleration and deceleration can be enforced to maintain passenger comfort. To handle uncertainty in the parameters, we utilize the tool of aCBFs to maintain and update estimates of these parameters. An aCBF that yields desirable results is defined as the following continuously differentiable function:
\begin{equation*}
    h_a(v,D) = \begin{cases} \alpha^2 & \quad \text{if } D-1.8v\geq \alpha \\ \alpha^2 - (D-1.8v-\alpha)^2 & \quad \text{if } D-1.8v < \alpha \end{cases}
\end{equation*}
for $\alpha>0$. This particular construction of $h_a$ is constant away from the safety boundary and diminishes to $0$ (quadratically to preserve differentiability) as the boundary is approached. In practice, this is to handle the fact that superlevel sets of the composite safety function $h$ are forward invariant. In regions where $h_a$ is constant, $\frac{\partial h_a}{\partial\mb{x}}$, and thus the update law in \eqref{eqn:csfupdatelaw}, is $\mb{0}$, thus making $\dot{h}=0$ as in the first equality in Table \ref{table:hdotcalc}.

\vspace{0.2cm}
\noindent {\bf aCBF-QP Controller:} 
A simple proportional controller on tracking error $v-v_d$ can be implemented and achieve good tracking performance, but is not necessarily safe. A CBF alone would not ensure the safety of this controller with model uncertainty, but treating the proportional controller as $k_d$ in \ref{eqn:safety_critical_qp} with an aCBF, safety can be achieved.

\vspace{0.2cm}
\noindent {\bf aCLF-aCBF-QP Controller:} 
Additionally, we can unify aCLF and aCBFs in a quadratic program based controller to receive the benefits of optimal and adaptive tracking while remaining safe. Separate estimates of the parameters are mainted for the aCLF and the aCBF, as the form of the update laws in \eqref{eqn:clfupdatelaw} and \eqref{eqn:csfupdatelaw} may not be simultaneously satisfiable for only one estimate of the parameters. The CLF in \cite{ames2014control} on the velocity tracking error $v-v_d$, given by $V_a=(v-v_d)^2$, also satisfies the aCLF condition \eqref{eqn:adaptiveCLFcondition}. Letting $\mb{x}=(v,z)$ and $\widehat{\bs{\theta}}$ and $\widehat{\bs{\psi}}$ be parameter estimates associated with the aCLF and aCBF, respectively, we formulate a QP-based controller:
\begin{align}
\label{eqn:aCLFaCSF_QP}
\MoveEqLeft[9]\mb{k}(\mb{x},\widehat{\bs{\theta}},\widehat{\bs{\psi}}) = \,\, \underset{\mb{u} \in \cal{U}}{\argmin}   \quad J( u) + c_V(\mb{x})\delta_V + c_p(\mb{x})\delta_p  \nonumber\\
\mathrm{s.t.} \,\,\,\,\,\, L_{\tilde{\mb{f}}_{clf}}V_{a}(\mb{x},\widehat{\bs{\theta}})&+L_{\mb{g}}V_{a}(\mb{x},\widehat{\bs{\theta}})u\le -\alpha_3(\Vert \mb{x}\Vert,\widehat{\bs{\theta}})+\delta_V\nonumber \\ L_{\tilde{\mb{f}}_{cbf}}h_{a}(\mb{x},\widehat{\bs{\psi}})&+L_{\mb{g}}h_{a}(\mb{x},\widehat{\bs{\psi}})u\geq 0 \nonumber \\ & \qquad\qquad\qquad\, u \leq u_{max}  +\delta_p \nonumber \\ &\qquad\qquad\qquad\, u \geq -u_{max}-\delta_p \nonumber \\ &\,\,\qquad\quad\,\,\,\,\, \delta_V,\delta_p \geq 0 \nonumber
\end{align}
with parameter updates for $\widehat{\bs{\theta}}$ and $\widehat{\bs{\psi}}$ as in \eqref{eqn:clfupdatelaw} and \eqref{eqn:csfupdatelaw}, respectively. $\delta_V$ and $\delta_p$ are relaxations to the optimization problem to ensure its feasibility, while safety is ensured. The functions $c_V$ and $c_p$ are Lipschitz continuous and are used to achieve smoothness. With initial parameter estimates $\begin{bmatrix}\hat{f}_0 & \hat{f}_1 & \hat{f}_2 \end{bmatrix}=10\begin{bmatrix}f_0^\star & f_1^\star & f_2^\star \end{bmatrix}$ (less friction than modeled), the results of this controller appear in Figure \ref{fig:accsimresults}.

We see that the proportional controller fails to keep the vehicle safe, but filtering it with the aCBF-QP keeps it safe (with $D\geq1.8v$ for all time) even with model uncertainty. A CLF-CBF controller with no adaptive elements fails to either track the desired velocity (with steady state error) or keep the vehicle safe. The CLF-aCBF controller keeps the vehicle safe but has steady state tracking error, while an aCLF-aCBF controller accurately tracks the desired velocity with no steady state error, and keeps the vehicle safe. 

\section{Conclusion}
We presented a novel approach for ensuring the safety of a system under a form of parametric uncertainty. This approach builds off the structure established with adaptive Control Lyapunov Functions, and highlights the differences that must be considered when ensuring the forward invariance of a specific set. Future work includes considering this framework within a batched-data framework, in which initial parametric uncertainty can be iteratively and episodically reduced to permit less conservative safe sets.


\gap
\bibliographystyle{plain}

\bibliography{main}


\end{document}